\documentclass[3p,times,procedia]{elsarticle}
\usepackage{nupha_ecrc}

\volume{00}

\firstpage{1}

\journalname{Nuclear Physics A}

\runauth{Wuppertal-Budapest collaboration}

\jid{nupha}

\jnltitlelogo{Nuclear Physics A}

\usepackage{amssymb}

\usepackage[figuresright]{rotating}

\begin{document}

\begin{frontmatter}



\dochead{}

\title{Towards the QCD phase diagram from analytical continuation}


\author[a]{R. Bellwied}
\author[b]{S. Bors\'anyi\footnote{speaker: borsanyi@uni-wuppertal.de}}
\author[b,c,d]{Z. Fodor}
\author[b]{J. G\"unther}
\author[c]{S. D. Katz}
\author[b]{A.~P\'asztor}
\author[a]{C.~Ratti}
\author[d]{K.~K.~Szab\'o}

\address[a]{Department of Physics, University of Houston, Houston, TX 77204, USA}
\address[b]{Department of Physics, Wuppertal University, Gaussstra\ss e 20, D-42119 Wuppertal, Germany}
\address[c]{Institute for Theoretical Physics, E\"otv\"os University, H-1117 Budapest, Hungary}
\address[d]{J\"ulich Supercomputing Centre, Forschungszentrum J\"ulich, D-52425
J\"ulich, Germany}

\begin{abstract}
We calculate the QCD cross-over temperature, the equation of state
and fluctuations of conserved charges at finite density
by analytical continuation from imaginary to real chemical potentials. Our
calculations are based on new continuum extrapolated lattice simulations using
the 4stout staggered actions with a lattice resolution up to $N_t=16$.  The
simulation parameters are tuned such that the strangeness neutrality is
maintained, as it is in heavy ion collisions.
\end{abstract}


\end{frontmatter}


\section{Introduction\label{sec:intro}}

An important goal of the heavy ion experiments at the Relativistic Heavy Ion
Collider (RHIC) at Brookhaven is to study the quark gluon plasma (QGP)
at various points in the QCD phase diagram. In the beam energy scan program
the collision energy and centrality determine the trajectory of the plasma
in the $T-\mu_B$ plane. This trajectory terminates in the chemical
freeze-out point, which corresponds to the instant of last inelastic
scattering. The collection of these points, the freeze-out curve gives an
experimental insight to the QCD transition between the hadron gas and the quark
gluon plasma phases.

Lattice QCD, on the other hand, works with equilibrium quantum field theory.
Using stochastic algorithms it can calculate bulk thermodynamics features,
like the order of the transition \cite{Aoki:2006we} transition temperature
\cite{Aoki:2006br,Aoki:2009sc,Borsanyi:2010bp,Bazavov:2011nk}, equation of
state \cite{Borsanyi:2010cj,Borsanyi:2013bia,Bazavov:2014pvz} as well as
fluctuations of conserved charges \cite{Borsanyi:2015axp} and a range of
correlation functions at any given finite temperature. Today it is possible to
run the simulations with the physical parameters of QCD and to perform a
continuum extrapolation, which is an essential step in lattice QCD. This
qualifies lattice results to be drawn as a reference point for the evaluation
of RHIC data \cite{Borsanyi:2014ewa}.

Lattice simulation algorithms for QCD with physical quark masses work at
vanishing baryochemical potential only. There are workarounds, however, that
enable us to extract finite-density information, nevertheless. Derivatives of
the equation of state, or the transition temperature can be calculated at zero
chemical potential, these are the Taylor coefficients for the extrapolation to
finite density. This method has an inherent limitation to small chemical
potentials. In praxis, the accessible range is likely to cover a large part of
the RHIC beam energy scan.

Higher Taylor coefficients are increasingly difficult to extract from $\mu_B=0$
simulations. A more efficient method to find these is coming from the fact that
lattice methods work also for imaginary chemical potentials. 
Analyticity connects the derivatives with respect to
$\textrm{Im}~\mu_B$ to the desired coefficients. The $\textrm{Im}~\mu_B$
dependence of e.g. $T_c(\mu_B)$ can be found from direct simulations at a set
of imaginary $\mu_B$ parameters \cite{DElia:2002gd,deForcrand:2002ci}.

\section{Strangeness neutrality\label{sec:neutrality}}

In this work we show the first results of our thermodynamics program
at imaginary chemical potentials. We use the 2nd generation (4stout)
ensembles of the Wuppertal-Budapest collaboration with the lattice resolutions
$N_t=10,12$ and 16. 

The range of available imaginary chemical potentials is limited to
$\textrm{Im}~\mu_B\in [0,\pi T]$ by the Roberge-Weiss symmetry. For
the actual simulations we select four values from this range: $\mu^{(j)}_B= i
j\pi T/8$ with $j=3,4,5,6$. In addition we use $j=0$ ensembles as
a reference.

For the strange quark chemical potential the popular choices include
the use of equal chemical potential for all quarks ($\mu_s=\mu_u=\mu_d$)
or the suppression of the strange quark chemical potential $\mu_s=0$.
Instead of these two options we use the physical strangeness neutrality
condition $\left\langle S\right\rangle=0$. This is an implicit constraint
on the simulation parameters, it requires careful tuning. For details
see Ref.~\cite{Bellwied:2015rza}.

\begin{figure}[t]
\centerline{\includegraphics[width=\hsize]{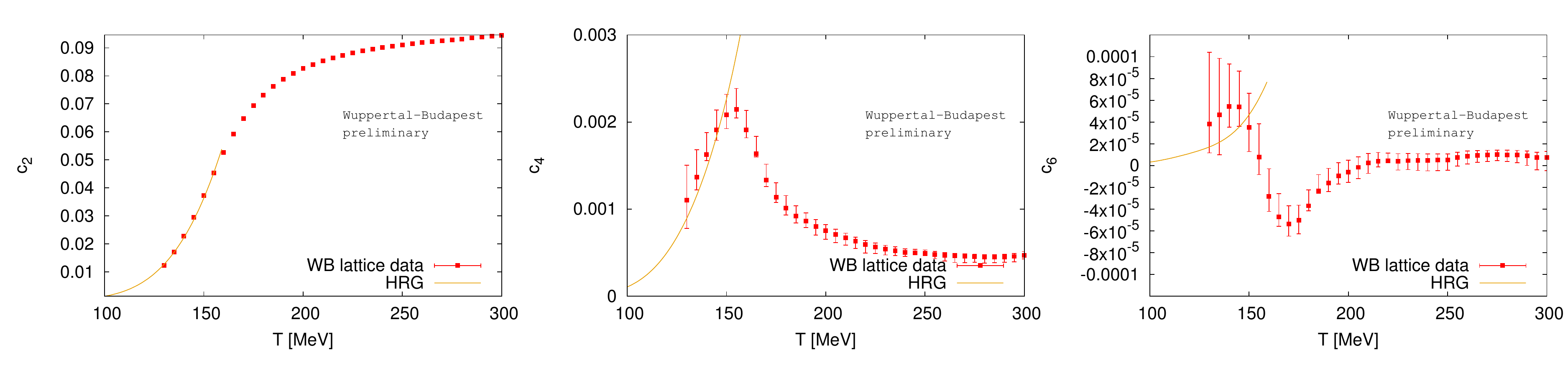}}
\caption{\label{fig:coeffs}
Taylor coefficients for the equation of state from imaginary $\mu_B$
simulations. The data are continuum extrapolated. 
The errors do not include the systematics of the fitting in $\mu_B$.
}
\end{figure}

\section{Transition line\label{sec:tcline}}

We calculated the transition temperature ($T_c$) for the four imaginary
chemical potentials in the continuum limit. We defined $T_c$ with three
observables as i) the peak of the renormalized chiral susceptibility normalized
to the fourth power of the pion mass, ii) the inflection point of the
renormalized chiral condensate, iii) the inflection point of the strangeness
susceptibility. The curvature $\kappa$ of the transition line in the QCD phase
diagram is defined by the equation
\begin{equation}
\frac{T_c(\mu_B)}{T_c(\mu_B=0)} = 1-\kappa \left(\frac{\mu_B}{T_c(\mu_B)}\right)^2+\mathcal{O}(\mu_B^4)\,.
\label{eq:kappa}
\end{equation}
Although different definitions are known to give slightly different
transition temperatures in the range around 155 MeV \cite{Borsanyi:2010bp},
they give remarkably consistent curvatures.
Our combined result is $\kappa=0.0149\pm 0.0021$ \cite{Bellwied:2015rza}.  The
transition line can be analytically continued, which we show in
Fig.~\ref{fig:phasediagram}. 
The central transition line corresponds to the inflection point of the
chiral condensate, its width shows the error of the extrapolation of the
inflection point. The increasing width is only partly coming from the
statistical errors. Instead of using Eq.~(\ref{eq:kappa}) we also
allowed a $\mu_B^4$ term in our fits. At large $\mu_B$ this
next-to-leading-order contribution drives the extrapolation error and signals
the end of the accessible $\mu_B$ range.

For recent continuum extrapolated lattice results, though without
implementing the strangeness neutrality condition, see
Refs.~\cite{Bonati:2015bha,Cea:2015cya}.

\section{Equation of state\label{sec:eos}}

The equation of state at finite density can be accessed through
the Taylor coefficients at $\mu_B=0$:
\begin{equation}
  \frac{p(\mu_B)}{T^4} = c_0(T)
+ c_2(T) \left(\frac{\mu_B}{T}\right)^2
+ c_4(T) \left(\frac{\mu_B}{T}\right)^4
+ c_6(T) \left(\frac{\mu_B}{T}\right)^6
+ \mathcal O(\mu_B^8)
\label{eq:eoscoeff}
\end{equation}
The first continuum result for $c_2$ was published in
Ref.~\cite{Borsanyi:2012cr}. In the physical point up to $c_4$ 
has recently been calculated, but without continuum extrapolation
\cite{Hegde:2014sta}.

The coefficients in Eq.~(\ref{eq:eoscoeff}) are defined such that
strangeness neutrality is implicitly assumed. In other words, $p/T^4$
is first expressed as function of $\mu_S,\mu_B$ and $T$, and evaluated
at $\mu_S(\mu_B,T)$ for which $\left\langle S\right\rangle=0$. Then
Taylor coefficients are defined then for each fixed $T$. Our results
also include a $\mu_Q$ to meet the actual setting in heavy ion
collisions, such that
$\left\langle Q\right\rangle=0.4\left\langle B\right\rangle$. 

Here we show results for the coefficients from imaginary $\mu_B$ simulations.
We fitted $c_2,\dots,c_6$ on the $\mu_B$-derivatives of $p/T^4$ for fixed
temperature, $c_0$ we determined earlier \cite{Borsanyi:2013bia}. The 
results are shown in Fig.~\ref{fig:coeffs}.

From the coefficients pressure, energy density, entropy and speed of
sound can be calculated at any (small) chemical potential. Here we
show one possible application: we calculate the trajectory of the
quark gluon plasma on the $T-\mu_B$ phase diagram. Since the
expansion of the plasma is adiabatic (constant entropy) and the net
conserved charges (e.g. baryon number) are constant in a closed system,
we can track the trajectory as the constant $s/n$ contours.

\begin{figure}[t]
\centerline{\includegraphics[width=12cm]{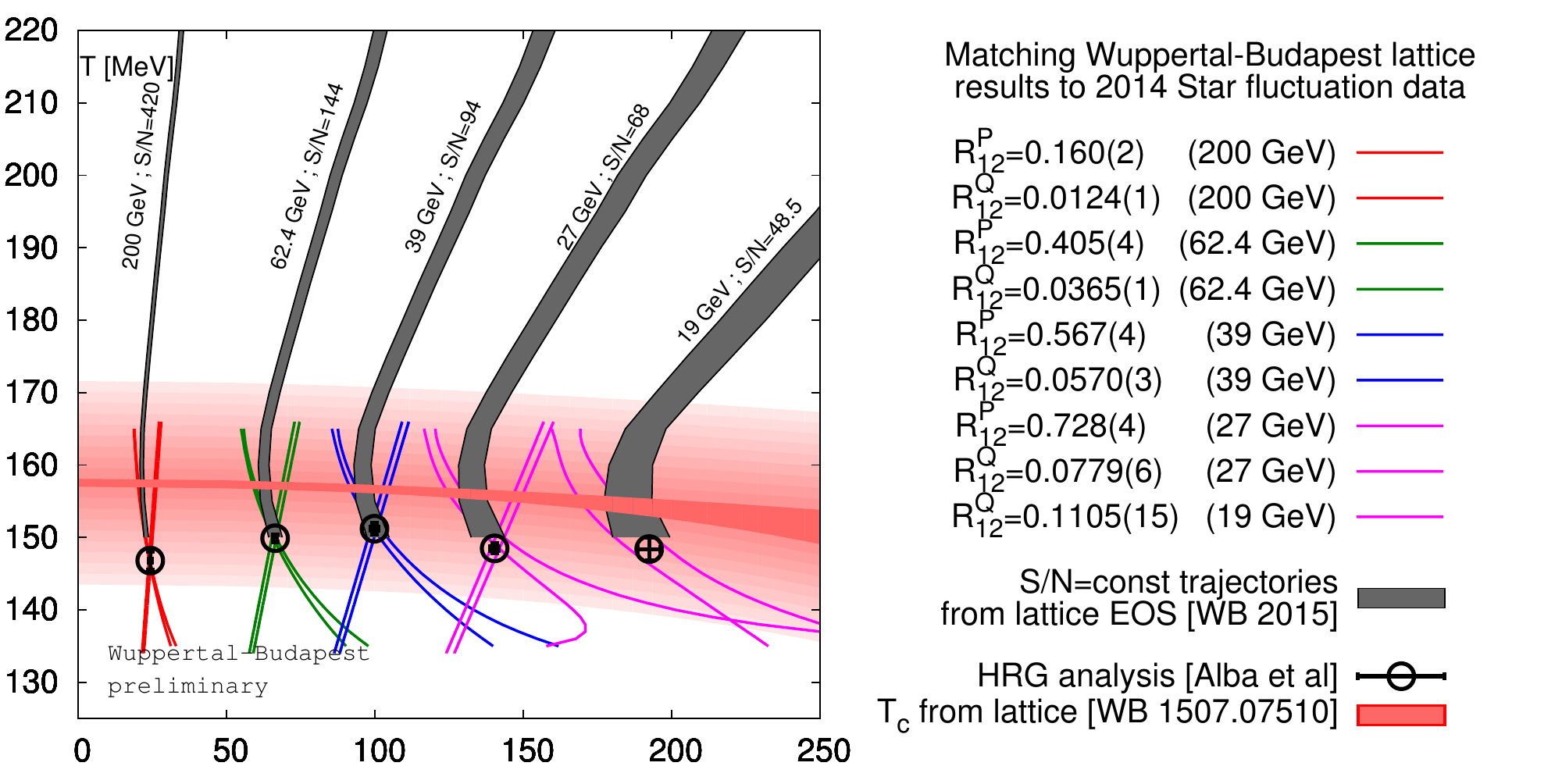}}
\caption{\label{fig:phasediagram}
The QCD phase diagram from analytical continuation. We used lattice simulations
with imaginary chemical potentials and extrapolated the transition temperature
(red band) to real chemical potentials. We also determined the equation of
state. Here we show the constant entropy/net baryon number contours that match
chemical freeze-out data. Finally, we show the contours for constant
mean/variance ratios of the net electric charge from lattice. We also show
the HRG prediction for the proton fluctuation ratios.
The contours that correspond to STAR data intersect in the freeze-out points of
\cite{Alba:2014eba}.
}
\end{figure}

For the central bin of each RHIC beam energy down to 19 GeV we find the
$s/n$ ratio in the freeze-out points located by the HRG-based analysis of
charge and proton fluctuations \cite{Alba:2014eba}.
Then we draw the entire contour in the phase diagram. We have checked
that the trajectory is consistent with the HRG prediction for all
collision energies near the freeze-out point. We show the contours and 
the transition line in Fig.~\ref{fig:phasediagram}. 

\section{Freeze-out curve\label{sec:freezeout}}

As an alternative to hadron yields, fluctuations of conserved
charges can also be used to find the freeze-out parameters, since lattice has
already calculated the equilibrium temperature dependence of many of the
fluctuation ratios
\cite{Bazavov:2012vg,Borsanyi:2013hza,Borsanyi:2014ewa}.
The direct comparison of the equilibrium ratios of lattice 
to experimental reality is not free from ambiguities
\cite{Bzdak:2012an,Skokov:2012ds}, the study of these goes beyond the scope of
this work.

Our results are summarized in Fig.~\ref{fig:phasediagram}. We calculated the
net electric charge mean/variance ratios for various imaginary chemical
potentials and extrapolated to finite $\mu_B$. In Fig.~\ref{fig:phasediagram}
we show the constant $M/\sigma$ contours matching the 2014 STAR data for
electric charge fluctuations \cite{Adamczyk:2014fia}. 
To avoid the use of baryon data from lattice as proton fluctuations
we simply used the HRG model to calculate the proton fluctuation ratio
and matched that to STAR data \cite{Adamczyk:2013dal}. 
If the comparison between lattice and experiment had
no additional systematics, the intersection points between these contours would
pin-point the freeze-out parameters.

Very recently, a similar study (but using $\mu_B=0$ ensembles) have been
published, where the $\kappa$ was extracted from fluctuation data
\cite{Bazavov:2015zja}. Interestingly the data seem to prefer a negative
freeze-out curvature, which is also true in our Fig.~\ref{fig:phasediagram} and
also in \cite{Alba:2014eba}. The full systematics of the comparison between
lattice and experiment is yet to be understood.

\noindent\textbf{Acknowledgements:} 
This project was supported by the Deutsche Forschungsgemeinschaft grant
SFB/TR55.  C. Ratti is supported by the National Science Foundation through
grant  number NSF  PHY-1513864.  S. D. Katz is funded by the "Lend\"ulet"
program of the Hungarian Academy of Sciences ((LP2012-44/2012).  An  award  of
computer time  was provided by the INCITE program.  This research used
resources of the Argonne Leadership Computing Facility, which is a DOE Office
of Science User Facility supported under Contract DE-AC02-06CH11357. The Gauss
Centre for Supercomputing (GCS) has provided computing time as a Large-Scale
Project on the supercomputer JUQUEEN. 





\bibliographystyle{elsarticle-num}
\bibliography{thermo}







\end{document}